\documentclass[prb,twocolumn,showpacs,preprintnumbers,amsmath,amssymb]{revtex4}
\usepackage{amsmath}
\usepackage{bm}
\usepackage{graphicx}
\begin{document}

\title{Coulomb interaction effects on the electronic structure \\
 of radial polarized excitons in nanorings}
\author{Z. Barticevic}
\author{ M. Pacheco}
\affiliation{Departamento de F\'{i}sica, Universidad T\'{e}cnica Federico
Santa Mar\'{i}a, \\ Casilla 110-V, Valpara\'{i}so, Chile,}
\author{J. Simonin}
\author{C. R. Proetto}
\affiliation {Centro At\'{o}mico Bariloche and Instituto Balseiro, \\
8400 S.C. de Bariloche, R\'{i}o Negro,  Argentina}
\date{july 2005}

\begin{abstract}
The electronic structure of radially polarized excitons in structured
nanorings is analyzed, with emphasis in the ground-state
properties and their dependence under applied magnetic fields
perpendicular to the ring plane. The electron-hole Coulomb
attraction has been treated rigorously, through numerical
diagonalization of the full exciton Hamiltonian in the non-interacting electron-hole pairs basis. Depending on the relative weight of
the kinetic energy and Coulomb contributions, the ground-state of
polarized excitons has ``extended" or ``localized"
features. In the first case, corresponding to small rings dominated by the kinetic energy, the ground-state shows Aharonov-Bohm (AB)  oscillations due to the individual orbits of the building particles of the exciton. In the localized regime, corresponding to large rings dominated by the Coulomb interaction, the only remaining AB oscillations are due to the magnetic flux trapped \textit{between} the electron and hole orbits. This dependence of the exciton, a neutral excitation, on the flux difference confirms this feature as a signature of Coulomb dominated polarized excitons. Analytical approximations are provided in both regimens, which accurate reproduce  the numerical results.
\end{abstract}
\maketitle
\section{Introduction}

Nanoscale semiconductor structures have been the subject of
numerous theoretical and experimental investigations in the last
few years. The effects of quantum confinement in these nanosystems
strongly modify their electronical and optical properties,
offering exciting possibilities for technological applications.
Among these, a particular class of structures with annular
geometry called nanorings are being intensively investigated
after the experimental observation of the
AB\cite{Aharonov} effect in small metallic
  rings\cite{Buttiker,Chandrasekhar,Mailly,Keyser,Tonomura}.
 With the strong development in the nanofabrication,  it is now
possible the formation of different types of  semiconductor
nanorings\cite{Lee}. This  gives us the exciting opportunity to
observe new quantum interference phenomena in magneto-optical
experiments\cite{Bayer,Lorke}. Several theoretical papers have
reported studies about the influence of the different
geometric-confinement parameters and the presence of impurities on
the spectrum in a semiconductor quantum ring in magnetic fields
\cite{Halonen,Barticevic,Bruno,Li}. The effects of an external
electric field on the AB oscillations in the energy
spectrum of quantum rings have been also reported\cite{Fuster}.
Most of the experimental work has been performed on charged
excitons in nanorings\cite{Bayer,Lorke,Warburton,Haft} and in
neutral excitons in type-II quantum dots\cite{Ribeiro}. The
possibility of the observation of the so-called ``optical" AB has
been an interesting and controversial subject in the recent years
\cite{Hu,Song,Climente,Romer}. It was predicted that the
polarization of a neutral exciton in a quantum ring can originate
a magnetic interference effect such that the ground-state of the
exciton alternates between states with zero (bright) and  nonzero (dark) angular momentum for increasing
magnetic field\cite{Ulloa,Govorov,Silva,Dias}. The finite
polarization of the exciton can be obtained by asymmetries in the
confinement potentials of the electron and holes or by means of a
uniform electric field applied in the ring plane\cite{Maslov}. In
the present paper we report a study of the effects of the Coulomb
interaction on the electronic structure of excitons in nanorings.
We consider radially polarized excitons and  we make a  detailed
analysis of the ground-state properties and its dependence with
magnetic fields applied perpendicular to the ring plane. We
include rigorously the electron-hole Coulomb interaction and
discuss different regimes of excitonic confinement. We also provided analytical approximations which are useful for semi-quantitative estimations in well-defined regimens.

\section{The model and method of solution}

The effective-mass excitonic Hamiltonian  in a quantum-ring
structure subject to an external magnetic field parallel to the
ring axis, which we take to be the $z$ axis, can be simplified
under some suppositions. In the first place, the electron and hole
coordinates along the $z$-direction may be ``frozen" at the same
in-plane value (say, $z_e = z_h =0$). This is consistent with the fact
that for all the semiconductor quantum rings produced by today's
semiconductor growth techniques, the confinement along the
$z$-direction (usually given by a compositional barrier) is much
stronger that the in-plane confinement. This gives rise to a
strong quantization along $z$. In the second place, the radial
displacements of the electron and hole may also be ``frozen" at
different radial coordinates, $ R_e $ and $ R_h $ respectively.
This is done by assuming that the effective self-consistent potentials,
for electron and for hole, have different radial positions for their
 respective minima\cite{jakak} and by realizing that the quantization in the radial direction
is usually stronger than in the azimuthal direction, for both of them.
These two approximations leads directly to\cite{Govorov}:
\begin{equation}
\hat{H}_{exc} (\theta_e , \theta_h) = \hat{H}_{exc}^{0}
(\theta_e, \theta_h) + U_c(\Delta \theta)\ \  ,
\end{equation}
where
\begin{eqnarray}
\hat{H}_{exc}^{(0)} (\theta_e , \theta_h)
= \frac{\hbar^2}{2m_e R_e^2}\left( - i
\frac{\partial}{\partial \theta_e } +  \frac{\phi_e}{\phi_0}
\right)^2 + \nonumber \\
\frac{\hbar^2}{2m_h R_h^2} \left(  i
\frac{\partial}{\partial \theta_h } + \frac{\phi_h}{\phi_0}
\right)^2 \ \  ,
\end{eqnarray}
is the sum of the electron and hole kinetic energies, and the Coulomb interaction is given by
\begin{equation}
U_c (\Delta \theta) = - \frac{e^2}{\varepsilon
(R_e^2 + R_h^2)^{1/2}} \frac{1}{[ 1- r  \cos (\Delta \theta)]^{1/2}} \ \ .
\end{equation}
In the above equations $ \Delta \theta = \theta_e - \theta_h $,
$m_e$ and $m_h$ are the electron and hole effective masses, $(R_e,
\theta_e)$ and $(R_h, \theta_h)$ are the radial and angular
electron and hole polar coordinates, $\phi_e = \pi B R_e^2$,
$\phi_h = \pi B R_h^2$ are the magnetic fluxes threading the
electron and hole rings, and $\phi_0 = ch/e$ is the flux quantum.
$U_c (\Delta \theta)$ describes the Coulomb attraction
between the electron and the hole, with $\varepsilon$ the
dielectric constant of the semiconductor ring material, and the
parameter $r=2 (R_e /R_h)/[1 + (R_e /R_h)^2 ]$ determines the
shape and the strength of the Coulomb interaction. For $ r \rightarrow 0 \ (R_h \gg R_e) $ the Coulomb potential as a function of $\Delta \theta $
is nearly flat, while for $ r \rightarrow 1 \ \ (R_h \simeq R_e)$,
the potential has a pronounced minimum at $\Delta \theta = 0$.

Even after these simplifications, $\hat{H}_{exc} (\theta_e,
\theta_h)$ is not exactly solvable, at least analytically, due
to the Coulomb interaction. In consequence, we have
used the following direct numerical strategy: diagonalization of
Eq.(1) in the non-interacting electron-hole pairs basis generated by the eigenstates of $\hat{H}_{exc}^0 (\theta_e, \theta_h)$,
\begin{equation}
\hat{H}_{exc}^0 (\theta_e, \theta_h)  \psi^0_{\ell_e,
\ell_h} (\theta_e, \theta_h) = E_{\ell_e, \ell_h}^{(0)}
\psi^{0}_{\ell_e, \ell_h} (\theta_e, \theta_h) \ \ \ \ ,
\end{equation}
where
\begin{equation}
 \psi^{0}_{\ell_e,
\ell_h} (\theta_e, \theta_h) = \frac{1}{(2 \pi)} \ \ e^{i
\theta_e \ell_e } e^{i \theta_h \ell_h }\ \ \ ,
\end{equation}
and
\begin{equation}
E^{(0)}_{\ell_e , \ell_h}(B) = \frac{\hbar^2}{2 m_e
R_{e}^{2}} \left( \ell_{e} + \frac{\phi_{e}}{\phi_{0}} \right)^{2}
+ \frac{\hbar^{2}}{2 m_{h} R_{h}^{2}}
\left( \ell_{h} - \frac{\phi_{h}}{\phi_{0}} \right)^{2} \ ,
\end{equation}
where $\ell_e$ and $\ell_h$ ( $ = 0, \pm 1, \pm 2 ,
...$ ) are the electron and hole angular momentum quantum numbers,
respectively. The non-interacting eigenstates of Eq.(5) can also
be written in terms of $ \Delta \theta $ and a new angular
variable, given by
\begin{equation}
\theta_0 = ( I_e \theta_e + I_h \theta_h ) / I \ ,
\end{equation}
here $I_{e} = m_e R_{e}^{2}$ , $I_{h}= m_{h} R_{h}^{2}$, and $I=
I_{e} + I_{h}$ are the electron, hole, and total moments of
inertia, respectively. $\theta_{0}$ is then a generalized angular
``center of mass" (CM) coordinate, and describes the translation
of the {\it{whole}} exciton around the ring, while $\Delta \theta$
describes the internal (relative) exciton dynamics. Replacing in
Eq.(5), we obtain,
\begin{equation}
\psi_{\ell_{e} , \ell_{h}}^{0} (\theta_{0} , \Delta \theta) =
\frac{1}{2 \pi} \, e^{i \theta_{0} (\ell_{e} + \ell_{h})} e^{i
\Delta \theta (\ell_{e} I_{e}  - \ell_{h} I_{h} ) / I}.
\end{equation}
The important point to note now is that as the Coulomb interaction
only depends on $\Delta \theta$, the total angular momentum of the
polarized exciton $L \equiv \ell_e + \ell_h$ remains a
good quantum number even in the interacting regime; this is a consequence of the
 azimuthal rotational symmetry of the structured rings. Thus, the Hamiltonian matrix
  generated in the basis of Eq.(5) is block-diagonal, with each block corresponding
  to a given $L\,(= 0, \pm 1, \pm2, ...)$. The
numerical diagonalization of each block provides the eigenvalues and eigenstates
of $\hat{H}_{exc}$, which we
denote by $E_{L,n}(B)$ and $\varphi_{L,n} (\theta_0,\Delta
\theta)$, respectively, with $n= 1,2,...$. To obtain accurate
exciton energies and wave-functions from the basis generated from
Eq.(5), we truncate the basis by choosing an adequate set of
quantum numbers $\ell_e$ and $\ell_h \ (= -\ell_e + L ) $ in each $ L $ sub-space.
This set is chosen starting from the couple  $\ell_e$, $\ell_h$ which  correspond
to the magnetic-field dependent non-interacting ground-state energy. The size of
the basis is chosen large enough such that the results do not depend on it.
The only non-trivial point of this calculation scheme is the numerical evaluation
of the matrix elements of $U_c(\Delta \theta)$ between the non-interacting
eigenstates of Eq.(5). While formally these matrix elements can be written
in term of the so-call toroidal functions\cite{Dias}, for its practical and
accurate evaluation we have found more convenient the direct numerical
integration over the one-dimensional variable $\Delta \theta$. Also,
and considering the easy of its direct numerical evaluation, we do not
 recommend its evaluation through the large-size ring approximation\cite{Dias},
  as we have found that it leads to some over-estimation of the exciton binding energy.

\section{Results}

All the numerical results to be discussed below were obtained with
material parameters appropriate for GaAs, that is: $m_{e} = 0.067\
m_{0}$, $m_{h}=0.268\  m_{0}$, and $\varepsilon = 12.5$, with
$m_{0}$ being the bare electron mass. The effective Bohr radius for the
 electron ($a_0^*$) is then equal to $98.7$ \AA. Also, we have assumed that
$R_{e} \leq R_{h}$.

Before proceed with the results a brief note on terminology is worth to be
discussed. In principle, the numerical diagonalization of the exciton Hamiltonian
 results in a large number of eigenvalues, for each structured ring and
 magnetic field value. Our analysis, however, will be mainly concentrated
 on the \textit{lowest} of these eigenvalues, the ground-state exciton.
 This ground-state exciton will be characterized in turn as belonging to
 the weak-interacting (WI) regime, or to the strong-interacting (SI) regime
  (see below). Excited exciton states will more properly considered as
  electron-hole pairs.

\subsection{Weak Interacting Regime}

In this regime, corresponding to structured rings of small size,
the dominant contributions to $\hat{H}_{exc} (\theta_{e},
\theta_{h})$ are the kinetic energy terms, with the Coulomb
interaction acting as a small modification to the non-interacting
results. Including accordingly $U_c (\Delta \theta)$ in a
perturbative way, we obtain
\begin{equation}
E_{\ell_{e}, \ell_{h}}^{(1)}(B) = E_{\ell_{e},
\ell_{h}}^{(0)}(B)
  - \frac{2
e^{2}}{\pi \varepsilon (R_{e} + R_{h})} \, K \left[ \frac{4 R_{e}
R_{h}}{(R_{e} + R_{h})^{2}} \right] \,\, ,
\end{equation}
with $K(x)$ being the complete elliptic integral of the first
kind\cite{Abramowitz}. The second term on the r.h.s. of Eq.(9) corresponds to
the matrix element $ U_c (m) \equiv \langle  \psi_{\ell_{e} - m ,
\ell_{h} + m }^{0}$ $|U_{c}|\psi_{\ell_{e} ,
\ell_{h}}^{0}\rangle$, with $ m = 0$. This diagonal matrix element is
the \textit{same} for all couples
of $(\ell_{e}, \ell_{h})$ non-interacting electron-hole quantum numbers.
Besides, in this regime, $ |U_c (0)| \gg |U_c (m \neq 0)| $, which supports
the perturbative expression Eq.(9).
According to this result, the exciton energy spectrum in the WI regime
is the same as the non-interacting spectrum, but shifted rigidly
by a negative constant.

\begin{figure}
\includegraphics[width=\columnwidth]{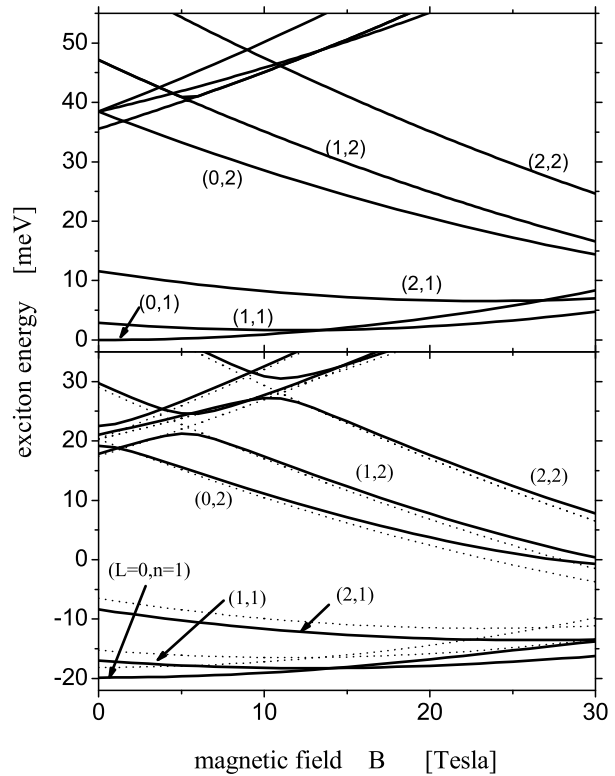}
\caption{Exciton energy versus magnetic field, for a ring with
$R_e=40\ \AA$, $R_h=70\ \AA$. Only states with total angular
momentum $L=0,1$ and $2$ are plotted, and for each of these,
the three lowest ones. Upper panel, non-interacting case.
Lower panel, interacting case: perturbative expression (dotted lines),
and numerical results (full lines).}
\label{fig1}
\end{figure}
We shown in Fig.\ref{fig1} the energy spectrum for a radial polarized exciton
(RPE) in
the WI regime, corresponding to a ring with $R_{e} = 40$\,\AA, and $R_{h} =
70$\,\AA. The top panel corresponds to the non-interacting
spectrum, the lower panel to the interacting spectrum calculated
exactly (numerically), and as given by the perturbative expression
of Eq.(9). As can be seen from the results in the lower panel, the
perturbative approximation nicely reproduces the main features of
the numerical result, shifting the non-interacting spectrum
towards negative energies by about $18\ meV$. It is interesting to
note that the approximation works better for excited than for low-lying states.
 This is
natural, as if the kinetic energy increases at constant Coulomb
correction, the accuracy of the perturbative approach should
increase. Beyond this simple first-order estimation are the several
anticrossings which appear in the numerical results when two
states with the same $L$ approach each other as function of $B$.

\subsection{From the weak to the strong interacting regime}

\begin{figure}
\includegraphics[width=\columnwidth]{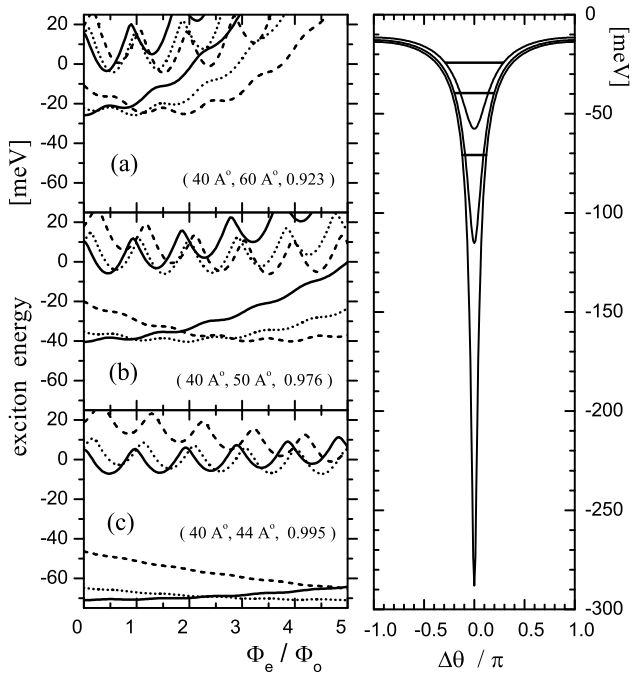}
\caption{Left panel, exciton energy spectrum versus electron magnetic flux,
  for three structured rings. Only the first low-lying
   states with $L = 0$ (full lines), $L=1$ (dotted lines), and $L=2$
   (dashed lines) are shown. Right panel: Coulomb attractive potential
   $U_{c} ( \Delta \theta )$   for the three structured rings of the left
   panel, the horizontal bars in them correspond to the lowest lying state (see text).}
\label{fig2}
\end{figure}

By approaching $R_{e}$ and $R_{h}$ to each other ($ r \rightarrow
1 $), the Coulomb attraction between the hole and the electron is
increasingly more important than the kinetic energy terms. We
shown this crossover in Fig.\ref{fig2}, where we display the RPE spectrum
for decreasing  values of $R_{h}$, keeping $R_{e}=40$\,\AA. The
more noticeable feature of these results is the progressive
appearance of a ``gap" among the low-lying and the excited
states, for each $L$ sub-space. Moving from top to bottom (left
panel), the modification of the spectrum consist  mainly in a
progressive ``deepening"  of the given $L$ low-lying state towards
negative energies, while the excited states remains at energies
close to zero. The right panel in Fig.\ref{fig2} corresponds to $U_{c}
(\Delta \theta)$, with the deeper one corresponding to $R_{h}= 44$\,\AA \,
($r = 0.995$), and the shallow one to $R_{h}= 60$\,\AA \, ($r = 0.923$). The
straight horizontal lines denote the position of the lowest-lying states
of the left panel (discounting the CM motion) for each size of the structured
 ring; once the kinetic energy of the CM motion has been subtracted the remaining
 energy becomes essentially independent of $L$ and $B$ (see below).
With this information at hand the meaning of the three split low-lying
states of Fig.2c is clear: they correspond to ``localized" exciton states,
whose wave-function (internal component) is strongly localized
around $|\Delta \theta | \simeq 0$. This must be contrasted with the
non-interacting exciton wave function of Eq.(8), whose internal
component is uniformly  distributed along its allowed values
($|\Delta \theta |\leq \pi$). Physically, this localization of the
ground-state exciton wave function is driven by the attractive
Coulomb interaction, which for $ r \rightarrow 1 $ is able of keep
the electron and hole as close as possible, loosing kinetic energy
but gaining Coulomb energy. An interesting feature of these
results is that as more localized is a state, less dependence on
the magnetic flux trapped by its individual components it shows.
This issue will be discussed in detail in the next sub-section.
We emphasize that the characterization of a state as ``extended" or
``localized" refers only to the internal component of the total exciton wave
function. The CM component is always extended, as corresponds to
a system with azimuthal rotational symmetry.

\subsection{Strong Interacting Regime}

\begin{figure}
\includegraphics[width=\columnwidth]{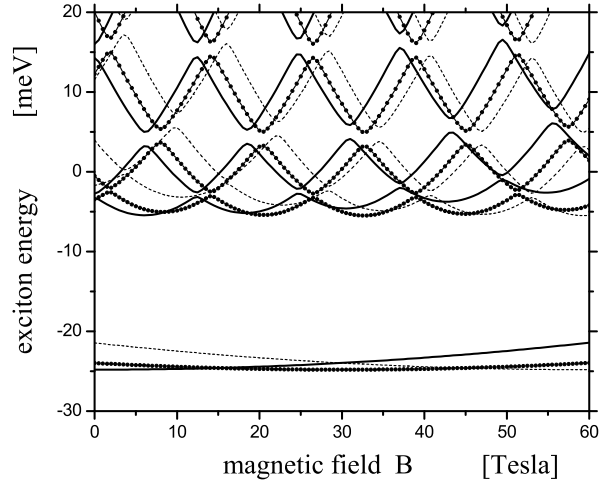}
  \caption{Exciton energy versus magnetic field for a ring with
  $R_e=100\ \AA$, $R_h=120\ \AA$. Only states with
  total angular momentum $L= 0$ (full lines), 1 (dotted lines),
  and 2 (dashed lines) are plotted.}
\label{fig3}
\end{figure}

\begin{figure}
\includegraphics[width=\columnwidth]{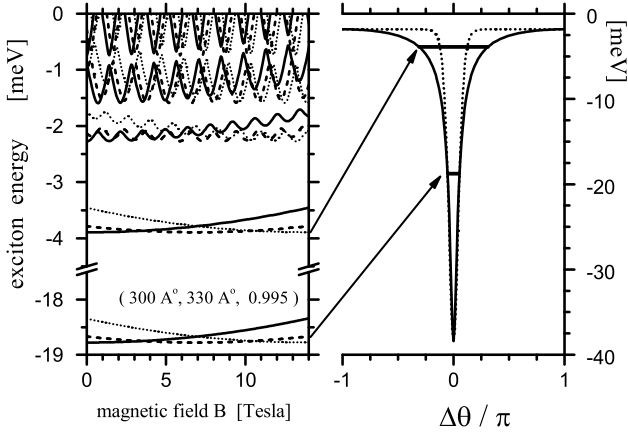}
\caption{Left panel, exciton energy spectrum versus magnetic field
   for a ring with $R_{e} =
  300$\,\AA, $R_{h} = 330$\,\AA. Only the first low-lying states
  with $L=0$ (full lines), $1$ (dashed lines), and $2$ (dotted
  lines) are shown. Right panel: Coulomb attractive potential
  $U_{c}(\Delta \theta)$ (solid line), $V_{c} (\Delta \theta )$
  (dashed line), and discrete energy levels (straight lines). Note
  the cut in the vertical axis of the left panel.}
\label{fig4}
\end{figure}

Increasing further the structured ring size, the system is driven
to the SI regime, where the electron and hole strongly interact.
Fig.\ref{fig3}, corresponding to a ring with $R_{e}= 100$\,\AA,
$R_{h}=120$\,\AA, displays clearly one set of localized $L$ ( $= 0$, $1$, and $2)$
 ground-states at negative energies, plus a bunch of closely energy spaced
 and strongly magnetic field dependent states at positive or close to
zero energies. Fig.\ref{fig4} corresponds to an even larger structured
ring, with the new feature of having \emph{two} localized
sets of states at negative energies, instead of one. The new localized
set corresponds to the first-excited state of each $L$.

An important feature of Figs.\ref{fig3}, \ref{fig4}, is the presence of a
characteristic energy (negative but close to zero), above of which
all states are extended. This energy is just $ U_c (\Delta
\theta = \pm \pi) = - \, e^2/\varepsilon (R_e + R_h)$,
corresponding to the minimum strength that the Coulomb potential can take
 in the constrained-ring geometry, and associated to the maximum possible
inter-particle distance . For the ring of Fig.2c,
$ U_{c} (\Delta \theta = \pm \pi) \cong -13.71$ $meV$, for the ring of
Fig.\ref{fig3} $ U_{c} (\Delta \theta = \pm \pi) \cong -5.23$ $meV$,
and about $-1.83$ $meV$ for the ring
of Fig.\ref{fig4}. Conversely, all states below that energy are localized
and their corresponding energies show a very weak dependence on
$L$ and $B$, as we will discuss latter.

To analyze properly the results of the SI regime, it is useful to
rewrite $\hat{H}_{exc} (\theta_{e} , \theta_{h})$ in term of the
variables $\theta_{0}, \Delta \theta$:
\begin{eqnarray}
\hat{H}_{exc} (\theta_{e} , \theta_{h}) &=& \hat{H}_{exc}
(\theta_{0} , \Delta \theta) = \hat{H}_{CM} (\theta_{0}) +
\hat{H}_{int} (\Delta \theta) \ \ ,
\end{eqnarray}
with
\begin{eqnarray}
 \hat{H}_{CM} (\theta_{0}) =\frac{\hbar^{2}}{2 I}  \left( - i
\frac{\partial}{\partial \theta_{0} } + \frac{\phi_{CM}}{\phi_{0}}
\right)^{2}\ \  ,
\end{eqnarray}
and
\begin{eqnarray}
 \hat{H}_{int} (\Delta \theta) =\frac{\hbar^{2}}{2 I_{int}} \left( - i
\frac{\partial}{\partial (\Delta \theta)} +
\frac{\phi_{int}}{\phi_{0}} \right)^{2} + U_{c} (  \Delta \theta
) \ \  .
\end{eqnarray}
The main achievement of this transformation is the {\it{exact}}
decoupling of the translational $(\theta_{0})$ and relative
coordinates $(\Delta \theta)$\cite{Govorov}. In equation above,
$\phi_{CM} = \pi (R_{e}^{2} - R_{h}^{2}) B$,  $I_{int} = I_{e}
I_{h} / I$, and $\phi_{int} = \pi I_{int} B/\mu$, with $\mu =
m_{e} m_{h} / (m_{e}+ m_{h})$.

Being $\hat{H}_{exc} (\theta_{0}, \Delta \theta)$ the sum of the
$CM$ and internal contributions, the corresponding eigenvalues are
given by
\begin{eqnarray}
 E_{L, n}(B) = \frac{\hbar^{2}}{2 I} \left( L +
 \frac{\phi_{CM}}{\phi_{0}} \right)^{2} + \varepsilon_{L,n} (B)\ \ ,
\end{eqnarray}
and
\begin{eqnarray}
 \varphi_{L,n} (\theta_{0}, \Delta \theta) = g_{L} (\theta_{0}) h_{L, n} (\Delta
 \theta) \,\, ,
\end{eqnarray}
with $g_{L} (\theta_{0})$ being eigenfunctions of
$\hat{H}_{CM}(\theta_{0})$ and $\varepsilon_{L,n} (B)$ and
$h_{L,n} (\Delta \theta)$ the eigenvalues and eigenfunctions of
$\hat{H}_{int} (\Delta
 \theta)$, respectively. While the solution of $\hat{H}_{CM}(\theta_{0})$
  is immediate, the solution of $\hat{H}_{int} (\Delta
 \theta)$ is not, due to its combined magnetic and Coulomb
 contributions, and the cumbersome cyclic boundary-conditions for
  its associated eigenfunctions. Some insight can be obtained, however,
  through
 the following gauge transformation of the internal component of
 the total exciton wave-function,
\begin{eqnarray}
 h_{L,n} (\Delta \theta) = e^{-i \Delta \theta \phi_{int}/\phi_{0} }
 f_{L,n} (\Delta \theta).
\end{eqnarray}
The boundary conditions for the total exciton wave-function are
easily established in term of the angular coordinates
$\theta_{e}$, $\theta_{h}: \varphi_{L,n} (\theta_{e}, \theta_{h})=
\varphi_{L,n} (\theta_{e} + 2 \pi, \theta_{h})= \varphi_{L,n}
(\theta_{e}, \theta_{h}+ 2 \pi)=  \varphi_{L,n} (\theta_{e}+ 2 \pi,
\theta_{h}+ 2 \pi)$. From these boundary conditions, it could be
easily derived an equivalent set of boundary conditions in terms
of $\theta_{0}$ and $\Delta \theta$: $\varphi_{L,n} (\theta_{0},
\Delta \theta) = \varphi_{L,n}(\theta_{0}+ 2 \pi I_{e} / I , \Delta
\theta + 2 \pi) = \varphi_{L,n}(\theta_{0}+ 2 \pi I_{h} / I, \Delta
\theta - 2 \pi)= \varphi_{L,n}(\theta_{0}+ 2 \pi \,, \, \Delta \theta
)$. Finally, using Eqs.(14) and (15), the boundary
conditions for the individual components of
$\varphi_{L,n}(\theta_{0}, \Delta \theta )$ are $g (\theta_{0})=
g(\theta_{0} + 2 \pi)$, and
\begin{subequations}
\begin{eqnarray}
f_{L,n}(\Delta \theta) =  \exp{[ 2 \pi i(\frac{I_{e} L}{I} + \frac{
\phi_{int}}{\phi_{0}})]}\  f_{L,n} (\Delta \theta + 2 \pi) ,
\\
f_{L,n}(\Delta \theta) = \exp{[ 2 \pi i(\frac{I_{h} L}{I} - \frac{
\phi_{int}}{\phi_{0}})] }\ f_{L,n} (\Delta \theta - 2 \pi) .
\end{eqnarray}
\end{subequations}
Replacing the anzatz of Eq.(15) in $\hat{H}_{int} (\Delta \theta)
h_{L,n}(\Delta \theta) = \varepsilon_{L,n}(B) \  h_{L,n} (\Delta
\theta)$, we derive an effective equation that defines $f_{L,n}(\Delta
\theta)$; this equation is:
\begin{equation}
- \frac{\hbar^{2}}{2 I_{int}} \frac{\partial^{2} f_{L,n} (\Delta
\theta)}{\partial(\Delta \theta)^{2}} + U_{c} (\Delta \theta)
f_{L,n} (\Delta \theta) = \varepsilon_{L,n} (B) f_{L,n} (\Delta
\theta)\,\,.
\end{equation}

The wave-function $f_{L,n}(\Delta \theta)$ satisfies then a
magnetic-field independent one-dimensional Schr\"oedinger-like
equation. The magnetic field dependence of $\varepsilon_{L,n}(B)$
is hidden now in the boundary conditions for $f_{L,n}(\Delta \theta)$.

Summarizing, $f_{L,n}(\Delta \theta)$ must satisfy the
Schr\"oedinger-like Eq.(17), plus the magnetic-field dependent
boundary conditions of Eq.(16). Now, and this is the whole
point, if the internal wave-function $f_{L,n} (\Delta \theta)$ is
strongly localized around $\Delta \theta \simeq 0$, it makes not difference if
we replace the complicated requirement of Eq.(16) by the
``isolated well" boundary condition $f_{L,n} (|\Delta \theta | \gg
1) \rightarrow 0$. Proceeding by this way, the eigenvalues
$\varepsilon_{L,n}$ become $L$ and magnetic-field independent, as
in this regime the internal Hamiltonian and the boundary condition
are \emph{both}, $L$ and magnetic-field independent. The approximation
works better the more localized is the state, and can be sought as
related to the tight-binding approximation employed in the
calculation of the band-structure of crystalline solids\cite{Kittel}. In this
last case, and for an atomic orbital strongly localized on the
scale of the lattice parameter, it makes no difference if one use
the rigorous boundary condition imposed by the Bloch theorem or
the ``isolated atom" boundary condition. In both cases, the
rigorous and the approximated calculation gives essentially the
same result: a discrete level just at the energy of the atomic
orbital of the isolated atom.

Accordingly, in this regime all the magnetic field dependence of
the exciton energy comes essentially from the CM contribution. Thus,
one can estimate the crossing points for exciton
states with different total angular momentum such as $L
\rightarrow L + M$ from the condition $E_{L,n}(B) = E_{L+M,n}(B)$,
\textit{neglecting} the $L$ and $B$ dependence of $\varepsilon_{L,n} (B)$.
Using Eq.(13), we obtain thus for the crossing magnetic fields,
\begin{equation}
B(L \rightarrow L+M) = \frac{\phi_{0}}{\pi(R_{h}^{2} - R_{e}^{2})}
\left( L + \frac{M}{2} \right) \ \ .
\end{equation}
It is interesting to note, in this regime, the strong sensitivity of the
 crossing magnetic-fields to the difference between $R_{e}$ and $R_{h}$.
 In particular, the optically active (bright) exciton with $L=0$ could be
 stabilized at larger magnetic fields by just moving to narrower structured
rings. Using Eq.(18) for the low-lying states of Figs. 3, 4, we
obtain for the ring of Fig.\ref{fig3} that $B(0 \rightarrow 1) \simeq
14.69$ T, and $B(0 \rightarrow 2) \simeq 29.92$ T. Proceeding in
the same way with the ring of Fig.\ref{fig4}, we obtain $B(0 \rightarrow 1)
\simeq 3.48$ T, $B(0 \rightarrow 2) \simeq 6.97$ T, and $B(1
\rightarrow 2) \simeq 10.45$ T. The good agreement between these
estimations and the exact (numerical) results, confirms the
hypothesis of the $L$ and magnetic-field independence of
$\varepsilon_{L,n}(B)$. It is also worth to be noted that the
crossing of the two sets of low-lying states $(n=1,2)$ in Fig.\ref{fig4}
takes place at the same crossing magnetic fields, as one expects
if Eq.(18) be valid.

The exciton energies $E_{L,n}(B)$ could be straight
forwardly calculated from Eq.(13),  once $\varepsilon_{L,n} (B)$
is known. For the localized regime discussed above we provide in
 Appendix A an analytic (approximated) solution to the problem posed by the
corresponding Eq.(17), which is useful for qualitative and semi-quantitative
estimations of $E_{L,n} (B)$. For instance, Eqs.(A.2) and (A.3) provide two
useful estimations of $\varepsilon_n$ (the $L$ and $B$ independent eigenvalues),
 and the number of bound localized states, respectively. It is interesting
  to note that this approximated analytical analysis \emph{always} predicted
  the existence of a localized state. This is in agreement with the output
  of the much elaborated numerical results.  It is  worth of be mentioned
  that the naive application of the harmonic approximation does not work
for the potential of Fig.\ref{fig4}. The reason for that is that $U_{c}
(\Delta \theta)$ is extremely deep and narrow (on the scale $- \pi
\leq \Delta \theta \leq \pi$). In consequence, approximating it by
an harmonic term in the bottom region, results in such a narrow
parabolic potential that the ground-state energy of the corresponding
harmonic oscillator (zero-point energy) is well above the ``continuum"
limit given by $U_{c} (\Delta \theta  = \pm \pi)$. In other words, the harmonic
approximation gives not bound-states for the ground-state exciton
for this structured ring, while the exact calculation yields two
bound-states.

\begin{figure}
\includegraphics[width=\columnwidth]{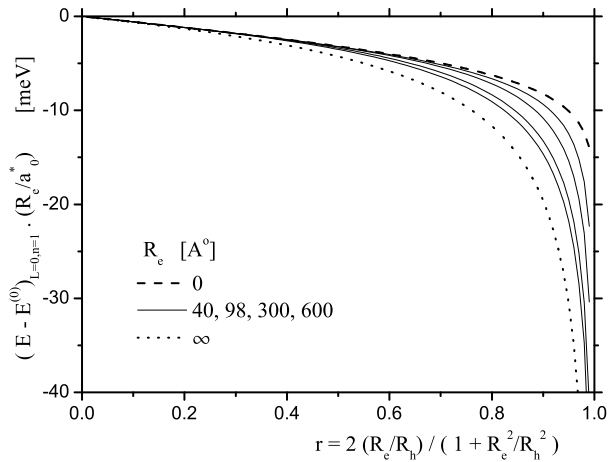}
\caption{Coulomb contribution to the ground-state energy, scaled by $R_e/ a_0^*$,
  at zero magnetic field and rings of several sizes. From top
  to bottom, $R_{e} \rightarrow 0$, $R_{e} = 40$ \AA, 98.7 \AA ($a_0^*$),
   $600$ \AA, and
  $R_{e} \rightarrow \infty$.}
\label{fig5}
\end{figure}
We display in Fig.\ref{fig5} the Coulomb contribution to the $L=0$, $n=1$
radial polarized exciton energy at zero magnetic field and for a set of
structured rings of different sizes, scaled by $R_e/a_0^*$. This
contribution has been numerically evaluated from the difference
between the interacting and non-interacting ground-state energies,
for each ring. The top curve corresponds to the Coulomb
perturbative correction of Eq.(9), which could be rewritten as
\begin{eqnarray}
\left( E_{\ell_{e} , \ell_{h}}^{(1)}(B) -E_{\ell_{e} ,
\ell_{h}}^{(0)}(B)\right) \left( \frac{R_{e}}{a_{0}^{\ast}}
\right) = \nonumber \\
 - \frac{2 e^{2}/a_{0}^{\ast}}{\pi \varepsilon \left( 1 +
R_{h}/R_{e} \right)} K \left[ \frac{4 R_{h}/R_{e}}{\left( 1 +
R_{h}/R_{e} \right)^{2}} \right].
\end{eqnarray}
Note that scaled this way the contributions for different
$R_{e}^{\prime}$s all collapses to a single curve. For $R_{h}$
approaching $R_{e}$ from above, the argument of the elliptic function
 tends to one and has a logarithmic divergence\cite{Abramowitz}. The curve at
the bottom corresponds to $U_{c} (\Delta \theta =0)$, scaled with
the same factor $R_{e}/ a_{0}^{\ast}$. Similarity to  the
contribution of Eq.(19), $R_{e} U_{c} (\Delta \theta=0)/
a_{0}^{\ast}$ collapses to a single curve for all values of
$R_{e}^{\prime}$s. In this case, for $R_{h}/R_{e} \rightarrow 1$,
the divergence is of the type $(R_{h} - R_{e})^{-1}$. The four
remaining intermediate curves, correspond from top to bottom to
$R_{e} = 40$\,\AA, $ R_{e}= a_{0}^{\ast}$, $R_{e} = 300$\,\AA, and
$R_{e} = 600$\,\AA. They have been obtained from the numerical
results. The result of Eq.(19) could be considered as giving the
limiting value of the Coulomb contribution in the WI regime of $R_{e}
\rightarrow 0$. This explains why the closer curve to this one is
that corresponding to the smallest considered rings, $R_{e} = 40$ \AA.
In a similar venue, the curve corresponding to $U_{c} (\Delta
\theta=0)$ could be considered as providing the limiting value
of the Coulomb contribution when $R_{e} \rightarrow \infty$, i.e. well
inside in the SI regime. In this extreme limit, the exciton behaves as a
\emph{classical} particle, its energy  given by the minimum of the
Coulomb potential at $ \Delta \theta = 0$. This explains why the
curves for increasing values of $R_{e}$ approach progressively
this limiting value.

\begin{figure}
\includegraphics[width=\columnwidth]{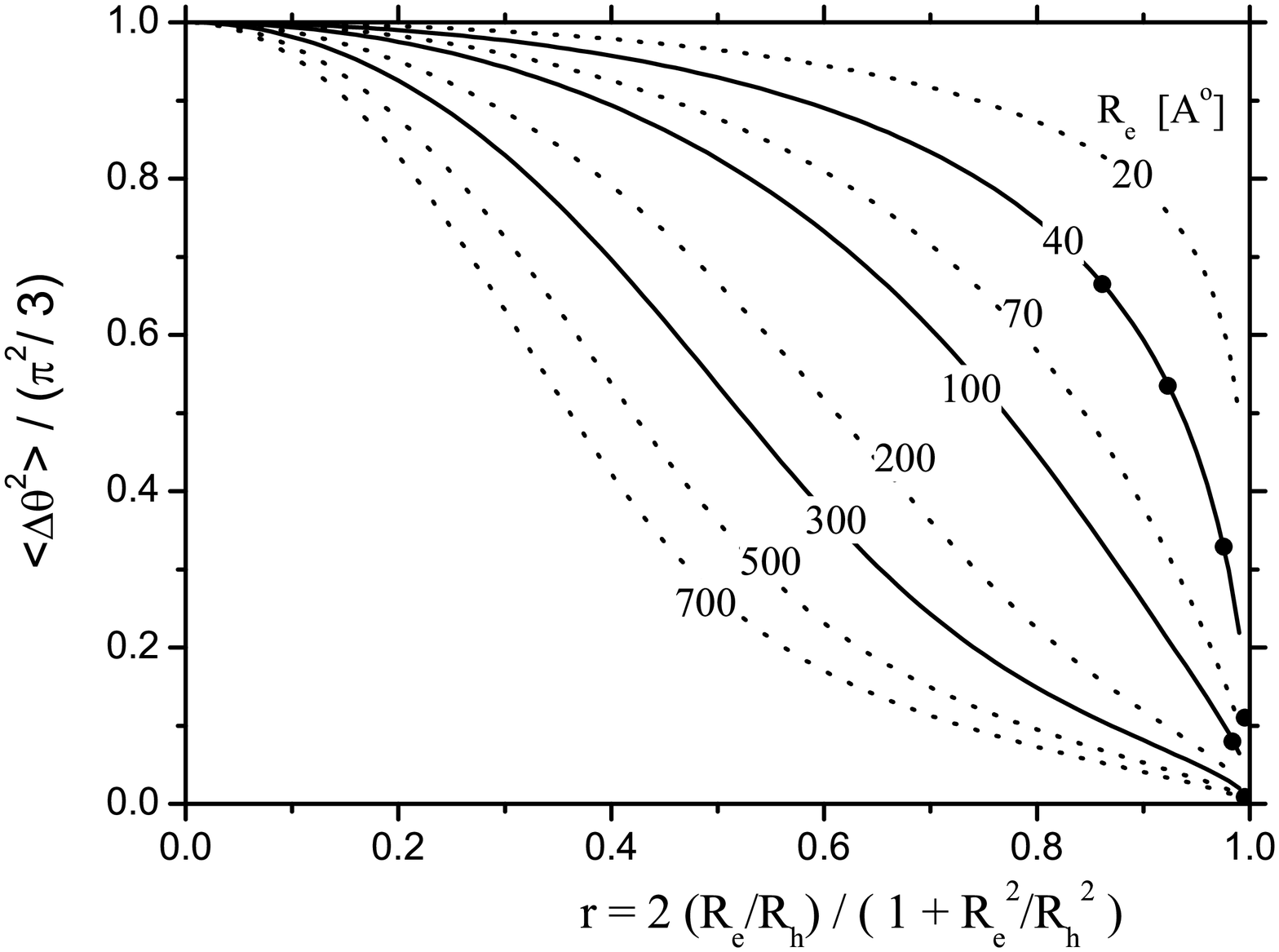}
\caption{Expectation value of $\Delta \theta^{2}$ in the ground-state
  at zero magnetic field, for several ring sizes. The dots on the curves
  with full lines correspond to the particular ring sizes presented in the
  previous figures.}
\label{fig6}
\end{figure}
As an accurate way of characterize the radial polarized
excitons in structured rings we have also calculated the
expectation value of $\Delta \theta^{2}$, in the ground-state of
the system at zero magnetic field. The results are presented in
Fig.\ref{fig6}. In general,
\begin{equation}
\langle (\Delta \theta)^{n} \rangle \equiv \int_{-\pi}^{\pi} d
(\Delta \theta)\  (\Delta \theta)^{n}\  | h (\Delta \theta) |^{2} \,\, ,
\end{equation}
with $h (\Delta \theta)$ the ground-state of $\hat{H}_{int}
(\Delta \theta)$. As a useful limit, we can evaluate Eq.(20) using
the non-interacting  zero-field ground-state eigenstate of
Eq.(14), which fulfills the normalization condition $|h (\Delta
\theta)|^{2}= 1/2 \pi$. Replacing this in Eq.(20), we obtain
\begin{equation}
\langle \Delta \theta ^{n} \rangle = \frac{\pi^{n+1}}{2\pi (n+1)}
\left[ 1 - (-1)^{n+1} \right] \ \  .
\end{equation}

This yields $\langle \Delta \theta \rangle =0$, and  $\langle
\Delta \theta ^{2} \rangle = \pi^{2}/3$. In consequence, we have
plotted $\langle \Delta \theta ^{2} \rangle / (\pi^{2} /3 )$ in
Fig. 6, versus $r$. When the value of this magnitude is close to
1 we can characterize the ground-state of the RPE  as extended.
Conversely, if $\langle \Delta \theta ^{2} \rangle / (\pi^{2} /3 )
\ll 1$, the ground-state is localized. The
discrete points on the curves corresponds to the structured rings
studied in this paper. The utility of this figure lies in
the fact that given an arbitrary ring, with the only information
of its geometrical dimensions, it is possible to obtain
immediately a qualitative characterization of its ground-state as
extended or localized. It is interesting to note that for intermediate
values of $r$ $(r \simeq 0.5)$, structured rings with the same
aspect ratio $R_{e}/R_{h}$ could be either in the WI regime
$(R_{e}=20\AA)$, or in the SI regime $(R_{e} = 700\AA)$. For small $(r
\simeq 0)$ or large $(r \simeq 1)$ values of $r$, all rings are
either in the WI  or in the SI regime, respectively.

\section{Conclusions}

We provide an accurate description of the electronic properties of
excitons in structured rings, concentrating mainly on the exciton
ground-states, and their response to magnetic fields applied
perpendicular to the ring plane. Our numerical method allows a
straightforward and precise calculation of ground-state energy
magnitudes for any size of the ring, magnetic field strength, and
material value parameters.

We have found that the ground-state of polarized excitons can be
well characterized in two extreme regimens: \emph{i)} The weak
interacting regime, where the electron and hole kinetic energies
are larger than the Coulomb interaction, and \emph{ii)} The strong
interacting regime, where the exciton ground-state properties are
dominated by the Coulomb electron-hole attraction.

For the weak interacting regime we have provided an analytical
approximation. According to this, the weakly interacting excitonic
spectrum could be obtained by shifting rigidly the non-interacting
spectrum by a negative constant. The constant depends only on the
size of the structured rings, but is state-independent. The
ground-state exciton WI shows discernible Aharonov-Bohm oscillations
with the magnetic field. In this regime it has been
predicted\cite{Govorov} the alternate of "bright" ($L=0$) and "dark"
 ($L\neq 0$) exciton ground-states as a function of the magnetic field.
However, for GaAs material parameters, we have found that the rigorous
inclusion of the Coulomb interaction removes such effect, and once the
$L=0$ ground-state crosses with the $L=1$ ground-state, it does
not become again the ground-state.

The ground-state in the strong interacting regime depends on
the magnetic field only through the phase accumulated by the CM angular
coordinate, which represents the translation as a whole of the
polarized exciton. On the other side, the relative angular
coordinate, that describes the internal dynamics of the polarized
exciton, remains essentially ``frozen" around zero. We have
provided also an analytical approximation for this regime.

We have found almost invariably, the simultaneous presence of
localized and extended states. This is somewhat similar to the
findings of Ref.[14], that have shown a similar excitonic spectrum
for the case of a ring with \emph{electric} and magnetic fields applied
along and perpendicular to the ring plane respectively, in
\emph{absence} of Coulomb interaction effects. The rotational and
internal structure of the exciton, however, are completely
different. In the situation of Ref.[14], the
localized states are induced by the applied electric field, that pushes
and localizes the electron and hole to opposite sides of the ring.
Using our notation and terminology, this kind of exciton has both, the rotational and
internal degree of freedom frozen, with $ \theta_{0} $ fixed by the
electric field direction, and $|\Delta \theta| \simeq \pi$.
In our case, the localized states correspond to a tightly-bound
electron-hole pair, whose relative coordinate is essentially ``frozen" at zero,
but whose center of mass coordinate rotates freely around the structured ring.

\renewcommand{\thesection}{\Alph{section}}
\setcounter{section}{0}
\section{Appendix}
\renewcommand{\theequation}{\thesection.\arabic{equation}}
\setcounter{equation}{0}

 We provide in this Appendix an analytic (approximated)
solution to the problem posed by the corresponding Eq.(17),
which is useful for qualitative estimations of $E_{L,n} (B)$ in
the localized regime. That is, neglecting the dependence of
$\varepsilon_{L,n} (B)$ on $L$ and $B$; these approximated
eigenvalues will be denoted by $\varepsilon_{n}$. With this aim, we
have found that $U_{c} (\Delta \theta)$ is well approximated by
\begin{equation}
U_{c} ( \Delta \theta )  \simeq V_{c} (  \Delta \theta  ) =  -
\frac{e^{2}}{\varepsilon} \frac{1}{|R_{e} - R_{h}|} -
\frac{V_{0}}{\cosh (\alpha  \Delta \theta)} \,\, ,
\end{equation}
where $V_{0} = e^{2} ( 1/ |R_{e} - R_{h}| - 1 / |R_{e} + R_{h}|
)/\varepsilon > 0$, and $\alpha$ is a dimensionless parameter to be
determined latter. The advantage of $V_{c} (\Delta \theta )$ over
$U_{c} (\Delta\theta )$ is that its exact analytical solution is
known\cite{Landau}. The eigenvalues associated with the bound
solutions are given by
\begin{equation}
\varepsilon_{n} = \frac{\alpha^{2} \hbar^{2} }{8 I_{int}} \left[ 1
-2 n +\left(1+ \frac{8 \mu V_{0}}{\alpha^{2} \hbar^{2}}
\right)^{1/2} \right]^{2} \ \ ,
\end{equation}
while the eigenfunctions are given in terms of the hypergeometric
function\cite{Landau}. It should be noted that Eq.(A2)
\emph{always} provided a bound state, corresponding to $n=1$. The
number of bound states is however finite, and given by the
condition
\begin{equation}
n_{max} < \frac{1}{2} \left[ 1+  \left(1+ \frac{8 \mu V_{0}}{\alpha^{2}
\hbar^{2}} \right)^{1/2} \right]  \ \ .
\end{equation}
In Eq.(A.2), everything is known, except $\alpha$. In our case,
as we have from our full diagonalization scheme the exact value of
$\varepsilon_{1}$, we have adopted the criterium of choose
$\alpha$ such that it reproduces exactly the numerical value of
$\varepsilon_{1}$, through Eq.(A2). This produces the value
$\alpha = 5.458$ for the structured ring of Fig. 4, and the $V_{c}
(\Delta \theta)$ shown as a dashed line in the right panel. This
``optimum" choice of $\alpha$ is reflected in the fact that $U_{c}
(\Delta \theta)$ and $V_{c} (\Delta \theta)$ are identical right
at the energy where $\varepsilon_{1}$ falls. We have checked, however,
 that the value of $\varepsilon_{1}$ is
not sensitive to the precise value of $\alpha$ and that any other
reasonable criteria for its determination (least-square fitting of
$U_{c} (\Delta \theta)$, etc.), works so well as our ``optimum"
fitting.

\section*{ACKNOWLEDGMENTS}

Z.B. and M.P. would like to thank financial support from Milenio
ICM P02-054-F and FONDECYT under grants 1020839 and 7020839. J.S.
and C.R.P. are fellows of CONICET, they thank partial financial
support from CONICET(Argentina) under grant PIP 02753/00.

\end{document}